\documentclass[
    ,final            
  ]
  {aipproc}
\usepackage{amsmath,amsthm}
\usepackage{amssymb}

\layoutstyle{8x11single}

\renewcommand{\(}{\left(}
\renewcommand{\)}{\right)}
\renewcommand{\[}{\left[}

\newcommand{\nn}{\nonumber}

\begin{document}

\begin{flushright} LU TP 15-04\\  January 2015 \end{flushright}

\title{Leading chiral logarithms for the nucleon mass}


\author{Alexey. A. Vladimirov}{
  address={Department of Astronomy and Theoretical Physics, Lund University,\\
S\"olvegatan 14A, SE 223 62 Lund, Sweden} }

\author{Johan Bijnens}{
  address={Department of Astronomy and Theoretical Physics, Lund University,\\
S\"olvegatan 14A, SE 223 62 Lund, Sweden} }

\begin{abstract}
We give a short introduction to the calculation of the leading chiral logarithms, and present the results of the recent evaluation of the LLog
series for the nucleon mass within the heavy baryon theory. The presented results are the first example of LLog calculation in the nucleon ChPT.
We also discuss some regularities observed in the leading logarithmical series for nucleon mass.  The talk has been presented at \textit{Quark
Confinement and Hadron Spectrum XI}.
\end{abstract}
\maketitle


\section{Introduction}

Chiral perturbation theory (ChPT) is an efficient tool for the evaluation of hadron observables at low energies (for a comprehensive
introduction to meson and nucleon ChPT, see \cite{Scherer:2005ri}). The predictions of ChPT are used in many branches of modern physics from
matching of lattice calculations (for a review see \cite{Aoki:2013ldr}), till the investigations of the nuclei properties (see e.g.
\cite{Epelbaum:2008ga}).

Nowadays, all practically interesting quantities has been calculated at two-loop order (for the recent status of meson ChPT see
\cite{Bijnens:2014lea}, also see the talk given by J.Bijnens \cite{Bijnens:2014qea}). However, the straightforward expansion to higher-orders of
ChPT is meaningless. The main point is the rapidly growing number of low-energy constants (LECs). Indeed, in the meson ChPT at order $p^6$ the
number of LECs is of order of hundred (depending on the number of active mesons)\cite{Bijnens:1999sh}. In the nucleon ChPT the number of fields
and invariant structures grows even faster, and this amount of LECs is reached already at order $p^4$ \cite{Fettes:2000gb}. Such an enormous
amount of LECs cannot be fixed in any reasonable way by a recent data.

Since the straightforward way to the higher precision theoretical predictions is closed, one should investigate other possibilities to improve
the chiral perturbation series. One of the promising approach is the evaluation of the leading logarithm (LLog) part of the chiral expansion.
Besides the possibility to improve the theoretical estimations for observables, the investigation of LLogs grants us a chance to understand the
mathematical structure of the theory at high orders of perturbative expansion. In contrast to renormalizable quantum field theories, where,
roughly speaking, LLog approximation consists in the powering of one-loop diagrams, LLogs in non-renormalizable theories are highly non-trivial.
The structure of LLogs in non-renormalizable theories resembles the structure of the whole perturbative expansion. Therefore, observation of any
regularities within LLog approximation is the reflection of general regularities.

In ChPT (as in any non-renormalizable theory) the evaluation of LLog coefficient of any given order implies the evaluation of diagrams at this
order. However, this can be done in a relatively efficient way. In this paper we give a short introduction to the calculation of the leading
chiral logarithms, and present the results of the recent evaluation of the LLog series for the nucleon mass: the first example of LLog
calculation in the nucleon ChPT.

\section{Leading logarithms in ChPT}

In ChPT the LLog coefficients can be calculated using solely one-loop calculations. This statement was proven in \cite{Buchler:2003vw}, and
relies on the fact that the LLog coefficient is proportional to the main ultraviolet (UV) divergency of a diagram. In its own turn, the leading
UV divergency of the diagram is composed from the individual divergencies of one-loop subgraphs. In the renormalizable theories, the individual
UV divergences of one-loop subgraphs are just multiplied on each other, while in ChPT the relation between subgraph's UV divergencies and the
leading divergency of a diagram are more involved.

There are several sources of complication. First of all, in ChPT every one-loop graph is UV divergent. As a result, all possible topologies of
diagrams contain LLog. This is the main complication point, because there are infinite number of one-loop graphs and all of them should be
evaluated in order to obtain the comprehensive LLog picture of ChPT. Second, since the loop-integrals in ChPT are dimensional, the individual
divergencies of graphs are entangled by the momentum structures. Despite its apparent simplicity, this is a serious problem, because it prevents
to derive any general relation between leading UV singularity of a graph and individual UV singularities of subgraphs without referring to the
particular expressions for graph.

In the bosonic ChPT this set of problems has been solved in refs.\cite{Buchler:2003vw,Kivel:2008mf,Kivel:2009az,Bijnens:2009zi,Bijnens:2010xg}.
The main idea of solution was to reconstruct the LLog coefficient by evaluation of the fixed chiral order solution of the system of
renormalization group equations on the amplitude and on LECs. Instead of giving the formal derivation of this method, which can be found in
\cite{Buchler:2003vw,Bijnens:2009zi,Bijnens:2014ila}, here we present an illustrative description of the route of calculation.

The base of the method is the formal solution of the renormalization group equation for an amplitude, which can be written in the form
\cite{Kivel:2008mf,Kivel:2009az}
\begin{eqnarray}
\label{defR} A(\{p\},\mu^2)=\exp\(\ln\(\frac{\mu^2}{\mu^2_0}\)\hat H\)A(\{p\},\mu^2_0),
\end{eqnarray}
where $\{p\}$ is a set of kinematical variables. The operator $\hat H$ acts on LECs and defined as
\begin{eqnarray}
\label{defH} \hat H= \int d\rho^2\sum_{n,i}\beta^{(n)}_i(\{c^{(m)}_j(\rho^2)\}) \frac{\delta}{\delta c_i^{(n)}(\rho^2)},
\end{eqnarray}
where $c_i^{(n)}$ are LEC of n'th chiral order, $\beta^{(n)}_i$ is the beta-function of the corresponded LEC and the variation over LEC is
defined as $\delta c^{(n)}(\mu^2)/\delta c_j^{(m)}(\rho^2)=\delta^{mn}\delta_{ij}\delta(\mu^2-\rho^2)$.

The action of the operator $\hat H$ on LEC replaces the LEC by its beta-function. On diagrammatic language, this procedure can be illustrated as
an insertion of the all possible (of given chirla order and of given number of external fields) one-loop graphs on the place of the vertex. The
crucial point of the approach is that the lowest order LECs (LECs of the second chiral order for meson ChPT, and LECs of the first and the
second chiral orders of nucleon ChPT) have zero beta-function. Therefore, the chain of one-loop insertions is finite.

\begin{figure}
  \includegraphics[height=.3\textheight]{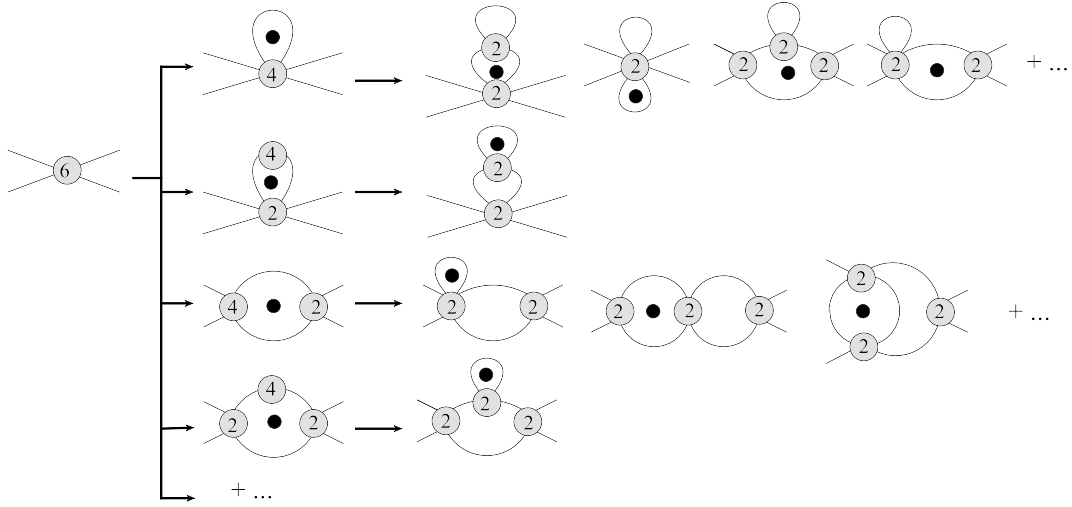}
  \caption{Illustration for the procedure of the LLog evaluation in the meson ChPT. The action of the operation $\hat H$ (plot as arrow) on a
  vertex effectively inserts the pole contribution of all possible one-loop subgraphs (inserted loop marked by a dot). The repetition of the operation inserts
  the next subgraph, and so on, until all the vertices of a graph are of the lowest order (so, no more insertion can be done). In this way one recovers the
  leading pole contribution of sum of all diagrams. Some diagrams appear in several brunches of the operational tree (see e.g. the first diagrams in the first and
  the second lines), this indicates the symmetry of a renormalization procedure and result to a correct multiplier for LLog coefficients. The numbers
  indicate the chiral order of vertices.}
\end{figure}

The procedure can be viewed as a graph tree, see fig.1.  The way from the root (the tree order diagrams) to branches (many-loop diagrams)
represents the particular chain of insertions of the one-loop graphs and illustrates the iterative solution of the renormalization group
equation (\ref{defR}). The consideration of the tree in the opposite way (from branches to the root) illustrates the traditional forest formula
for the renormalization of a diagram. In this way every step gives a subtraction of a subgraph. One can see that the same multi-loop diagrams
appear on the tree several times, this is a reflection of the multiple ways of the subgraph subtractions.

Important to mention that at every chiral order there are vertices of all possible momentum and isotopic structures whose non-trivially relate
to each other in beta-functions. Therefore, the chain of insertions (or subtractions, depending on the point of view) precisely restores the
leading UV singularity of a graph through the known UV singularities of one-loop subgraphs. Knowing the leading UV divergency, one easily
restores the LLog coefficient.

The described procedure can be used for the evaluation of the renormalization group logarithms only, i.e. the logarithms of the renormalization
scale $\mu$. The logarithms of momenta and masses should be considered as a finite parts, and do not participate in the logarithmical counting.
However, the procedure can be generalized on of any order logarithms by increasing the loop-order of beta-functions. For example, to obtain the
next-to-LLog coefficient one should use two-loop beta-functions, on the graph-tree that would correspond to the insertion of two-loop graphs in
one turn.

In some particular models the procedure can be simplified. The great example is the ChPT with massless fields. In this case tad-pole diagrams
are zero and many branches of the graph-tree lead to zeros (e.g. in fig.1 only the last two diagrams of the third line are non-zero). In this
case the infinite set of renormalization group equations can be presented as the closed recursive equation \cite{Kivel:2008mf,Kivel:2009az}. The
recursive equation has the straightforward interpretation as the LLog solution of the joined system of standard restrictions on the pion
amplitudes, namely, unitarity, analyticity and crossing symmetry \cite{Koschinski:2010mr}. Such a recursive equations are the convenient tool
for the investigation of effective field theories of a very general structure \cite{Polyakov:2010pt}.

The described method has been used for the calculation of LLog series for the pion mass, form factors, and scattering lengths
\cite{Bijnens:2009zi,Bijnens:2010xg,Bijnens:2012hf,Bijnens:2013yca}. The calculation can be fully automatized and then the maximum order of
calculation is restricted by the computing time. The typical order of calculation is the sixth power of logarithms, which corresponds to the
evaluation of six-loop diagrams. For example, the mass of pion in ChPT with two active flavors at this order  reads \cite{Bijnens:2012hf}
\begin{eqnarray}
\label{2:mPhys} m^2_\text{phys}=m^2\(
  1-\frac{1}{2}L+\frac{17}{8}L^2-\frac{103}{24}L^3+\frac{24367}{1152}L^4
  -\frac{8821}{144}L^5+\frac{1922964667}{6220800}L^6+\cdots\),
\end{eqnarray}
where
\begin{eqnarray}
\label{2:L_def} L=\frac{m^2}{(4\pi F)^2}\log\(\frac{\mu^2}{m^2}\).
\end{eqnarray}

The LLog contribution to the scattering amplitudes is not significant, e.g. see estimations in \cite{Bijnens:2010xg,Bijnens:2012hf}. Numerically
higher orders of LLog series give a tiny correction to the first terms and the series is very fast convergent. Anyway, the difference between
LLog and next-to-LLog of characteristic energy is much smaller then the difference between the powers of same energy. Therefore, the direct
practical application of LLogs is unreasonable. However, in the case of the chiral expansion for parton distributions at small-$x$ ($x\sim
\frac{m_\pi^2}{(4\pi F_\pi)^2}\sim 10^{-2}$) the powers of energies effectively cancel by powers of $x$, and LLog approximation is the dominant
one \cite{Kivel:2007jj,Moiseeva:2012zi}. That leads to the observable effect of the LLog resummation in ChPT
\cite{Perevalova:2011qi,Moiseeva:2013qoa}.

\section{LLogs for the physical mass of nucleon}
\label{sec:resultmass}

\begin{figure}
  \includegraphics[height=.35\textheight]{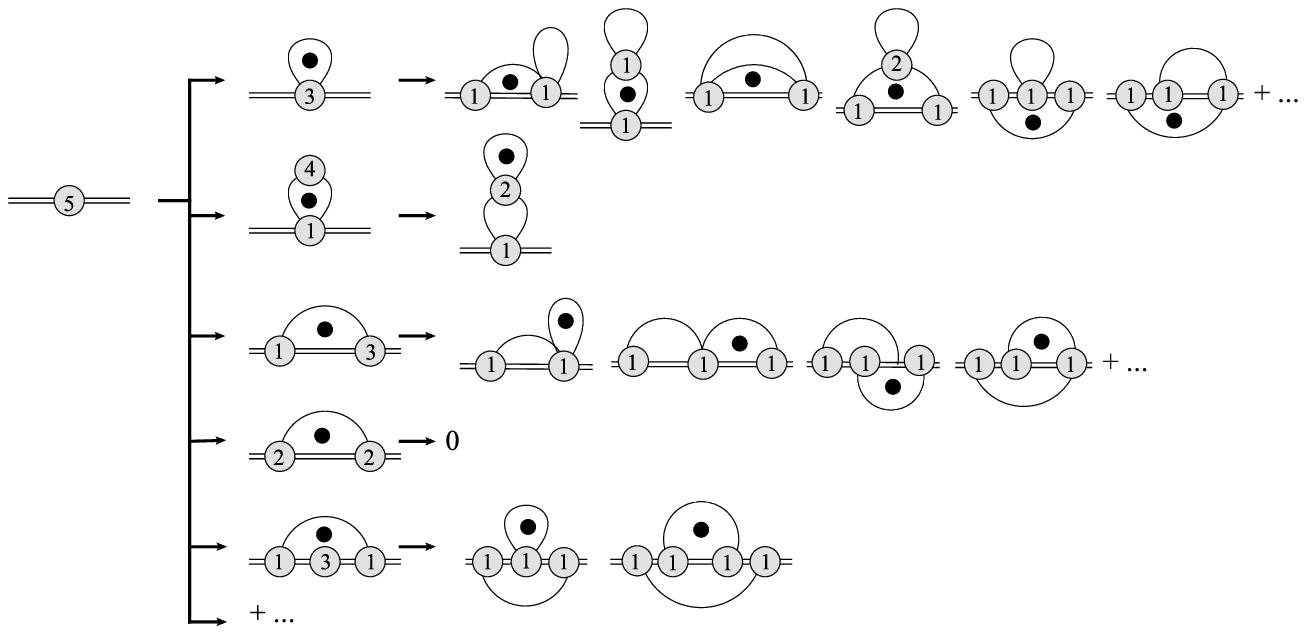}
  \caption{Illustration for the procedure of the LLog evaluation in the nucleon ChPT (see also capture of fig.1). In contrast to meson ChPT there are two sets of
  lowest order vertices, of even and of odd chiral orders. The sum of any two even-chiral-order-vertices has the same chiral order, as two
  odd-chiral-order-vertices joined by loop. Therefore, diagrams with more then one even-chiral-order-vertex do not contribute to LLog. On the operational tree
  the branches with such insertions result to zero (see the third line).}
\end{figure}

The generalization of the scheme on the theories with baryons is straightforward. There are only several complication points that should be
specially cared.

The first complication is the hierarchy problem of nucleon-pion system, which is the standard problem of the nucleon ChPT. Nowadays several
solution of the problem is known. The most popular are the infrared renormalization scheme \cite{Becher:1999he}, and the extended-on-mass-shell
renormalization scheme \cite{Fuchs:2003qc}. Both these schemes relies on the additional subtractions of the scale-violating terms, as a part of
the renormalization procedure. Since the additional (finite) subtraction do not spoil the renormalization, these schemes can be easily
implemented into LLog calculations. However, for the first try we found convenient to use somewhat old-fashion heavy baryon ChPT (HBChPT), see
e.g.\cite{Bernard:1992qa,Bernard:1995dp}. The advantage of HBChPT for LLog calculation is that no extra subtractions should be done, by the
price of violation of Lorentz invariance on the intermediate steps.

The second complication is that in HBChPT (as in any ChPT with nucleon) there are two lowest order Lagrangians. The first chiral order
Lagrangian, neglecting terms with external fields, reads
\begin{eqnarray}\label{3:L0}
\mathcal{L}^{(0)}_{N\pi}=\bar N\(i v^\mu D_\mu+g_A S^\mu u_\mu\)N,
\end{eqnarray}
where $g_A$ is the axial coupling constant, $S^\mu$ is a spin vector, $v^\mu$ is the nucleon velocity vector, $u$ is the matrix of pion fields,
and the field combinations read
\begin{eqnarray}
D_\mu&=&\partial_\mu+\Gamma_\mu, \qquad u_\mu=i\(u^\dagger \partial_\mu u-u\partial_\mu u^\dagger\), \qquad \Gamma_\mu= \frac{1}{2}\(u^\dagger
\partial_\mu u+u \partial_\mu u^\dagger\).
\end{eqnarray}
The second order Lagrangian is sensitive to redefinitions of the nucleon field, and in the most standard form reads
\cite{Bernard:1992qa,Bernard:1995dp}
\begin{eqnarray}
\label{3:BKKM} \mathcal{L}_{\pi N}^{(1)}&=&\bar N_v\Big[
 \frac{(v\cdot D)^2-D\cdot D-ig_A \{S\cdot D,v\cdot u\}}{2M}
 +c_1\text{tr}\(\chi_+\)+\(c_2-\frac{g_A^2}{8M}\)(v\cdot u)^2
 \nn\\&&
 \hspace*{1cm}
  +c_3 u\cdot u+\(c_4+\frac{1}{4M}\)i\epsilon^{\mu\nu\rho\sigma}u_\mu u_\nu v_\rho
          S_\sigma \Big]N_v,
\end{eqnarray}
where $M$ is the nucleon mass. The addition of the vertices with odd chiral counting does not spoil the general procedure of LLog computation,
but leads to extra algebra, see details in \cite{Bijnens:2014ila}. Indeed, since every loop increases the chiral counting of a diagrams by at
least two, the LLog of power $n$ is collected from the diagrams of two different scales, $(2n)$ and $(2n+1)$.

The third complication is the practical one. Due to the presence of the external vector $v^\mu$, complicated spin structure, as well as,
presence of two initial Lagrangians the calculation within HBChPT is dramatically larger in comparison to the meson ChPT. It is reflected in
every aspect of calculation: the number of diagrams, higher order vertices and the length of the counter terms. These quantities grow rapidly
with chiral order. For example, in order to calculate the four-loop LLog coefficient for nucleon mass one needs to evaluate nearly $10^4$
one-loop diagrams.

The main route of the calculation remains the meson calculation: one should evaluate one-loop beta-function and insert them into amplitude by
means of procedure (\ref{defR}). In order to obtain the correction to the nucleon mass we evaluate the nucleon propagator, an example of
diagram-insertion-chain is shown in fig.2. One can see that the number of diagrams and their topologies is significantly larger, in comparison
to the meson calculation (see fig.1). However, one can also see some minor simplifications, e.g. the fourth line in fig.2 is ended by zero. It
is the reflection of the fact that Lagrangians (\ref{3:L0}) and (\ref{3:BKKM}) have not counter terms. Therefore, the graphs with too many
vertices from $\mathcal{L}_{\pi N}^{(1)}$ have larger chiral counting but do not produce LLogs. Taking into account this fact significantly
reduce the number of diagrams to consider.

We have performed the calculation of the nucleon propagator and evaluate the nucleon physical mass. The procedure of LLog evaluation has been
automatized using the computation system FORM \cite{FORM}. Within a reasonable computing time we have calculated the LLog coefficient and the
non-analytical in quark mass LLog term up to the fourth power. The results are presented in the form:
\begin{eqnarray}
\label{mainresult}
M_{\text{phys}}&=&M+k_2 \frac{m^2}{M}+k_3 \frac{\pi m^3}{(4\pi F)^2}
  +k_4 \frac{m^4}{(4\pi F)^2 M}\log\(\frac{\mu^2}{m^2}\)
  +k_5 \frac{\pi m^5}{(4\pi F)^4}\log\(\frac{\mu^2}{m^2}\)+\cdots
\nn\\
&=& M+\frac{m^2}{M}\sum_{n=1}^\infty k_{2n} L^{n-1} +\pi m\frac{m^2}{(4\pi F)^2}\sum_{n=1}^\infty k_{2n+1} L^{n-1},
\end{eqnarray}
where $L$ is defined in (\ref{2:L_def}). The coefficients up to $k_{11}$ are presented in the table 1. This corresponds to the four-loop
calculation of LLog and five-loop calculation for the terms non-analytical in quark masses.

A very strong check of the calculation is performed by parallel calculation with the different parameterizations of the Lagrangians.
Additionally, the coefficients up to $k_6$ agree with known results. The one-loop coefficients $k_{3,4}$ are well known, see e.g.
\cite{Bernard:1995dp}. The two-loop coefficient $k_{5}$ was first derived in \cite{McGovern:1998tm}. The two-loop coefficients $k_6$ and $k_5$
are known from the full two-loop calculation for the nucleon mass performed in the EOMS scheme \cite{Schindler:2007dr}.

\begin{table}[tb!]
\centering
\begin{tabular}{|c|l||c|l|}
\hline \rule{0ex}{2.5ex} $k_2$ & $-4c_1 M$ & $r_2$ & $-4c_1 M$
\\[1mm]\hline
\rule{0ex}{2.5ex}$k_3$ & $-\frac{3}{2}g_A^2$ & $r_3$ & $-\frac{3}{2}g_A^2$
\\[1mm]\hline
\rule{0ex}{2.5ex}$k_4$ & $\frac{3}{4}\(g_A^2+(c_2+4c_3-4c_1)M\)-3c_1M $ & $r_4$ & $\frac{3}{4}\(g_A^2+(c_2+4c_3-4c_1)M\)-5c_1M $
\\[1mm]\hline
\rule{0ex}{2.7ex}$k_5$ & $\frac{3g_A^2}{8}\(3-16 g_A^2\)$ & $r_5$ & $-6 g_A^4 $
\\[1mm]\hline
\rule{0ex}{2.5ex}$k_6$ & $-\frac{3}{4}\(g_A^2+(c_2+4c_3-4c_1)M\)+\frac{3}{2}c_1M $ & $r_6$ & $5c_1M $
\\[1mm]\hline
\rule{0ex}{2.9ex}$k_7$ & $g_A^2\(-18 g_A^4+\frac{35 g_A^2}{4}-\frac{443}{64}\)$ & $r_7$ & $\frac{g_A^2}{4}\(-8+5g_A^2-72 g_A^4 \)$
\\[2mm]\hline
\rule{0ex}{2.5ex}$k_8$ & $\frac{27}{8}\(g_A^2+(c_2+4c_3-4 c_1)M\)-\frac{9}{2}c_1M $ & $r_8$ & $\frac{25}{3}c_1M $
\\[1mm]\hline
\rule{0ex}{2.9ex}$k_9$ & $\frac{g_A^2}{3}\(-116 g_A^6+\frac{2537 g_A^4}{20}-\frac{3569 g_A^2}{24}+\frac{55609}{1280}\)$ & $r_9$ &
$\frac{g_A^2}{3}\(-116 g_A^6+\frac{647 g_A^4}{20}-\frac{457 g_A^2}{12}+\frac{17}{40}\)$
\\[2mm] \hline
\rule{0ex}{2.5ex}$k_{10}$ & $-\frac{257}{32}\(g_A^2+(c_2+4c_3-4c_1)M\)+\frac{257}{32}c_1M $ & $r_{10}$ & $\frac{725}{36}c_1M $
\\[1mm]\hline
\rule{0ex}{2.9ex}$k_{11}$ & $\substack{\textstyle{\frac{g_A^2}{2}\Big(-95 g_A^8+\frac{5187407 g_A^6}{20160}-\frac{449039
   g_A^4}{945}+\frac{16733923 g_A^2}{60480}}\\ \textstyle{~~~~~~~~~~~~~~~~~~~~~~~~~~~~~~~~~~~~~~~~~~~-\frac{298785521}{1935360}\Big)}}$ & $r_{11}$ &
    $\substack{\textstyle\frac{g_A^2}{2}\Big(-95 g_A^8+\frac{1679567 g_A^6}{20160}-\frac{451799 g_A^4}{3780}+\frac{320557
   g_A^2}{15120}\\\textstyle ~~~~~~~~~~~~~~~~~~~~~~~~~~~~~~~~~~~~~~~~~~~~~~~~~-\frac{896467}{60480}\Big)}$
\\[2mm]
\hline \rule{0ex}{2.5ex}$k_{12}$(*) & $\frac{115}{3}\(g_A^2+(c_2+4c_3-4c_1)M\)-\frac{92}{3}c_1M $ & $r_{12}$(*) & $\frac{175}{4}c_1M $
\\[1mm]\hline
\rule{0ex}{2.5ex}$k_{14}$(**) & $-\frac{186515}{1536}\(g_A^2+(c_2+4c_3-4c_1)M\)+\frac{186515}{2304}c_1M $ & $r_{14}$(**) &
$\frac{4153903}{24300}c_1M $
\\[1mm]\hline
\rule{0ex}{2.5ex}$k_{16}$(**) & $\frac{153149887}{259200}\(g_A^2+(c_2+4c_3-4c_1)M\)-\frac{153149887}{453600}c_1M $ & &
\\[1mm]\hline
\end{tabular}
\caption{The coefficients $k_i$ and $r_i$ of the LLog expansion of the nucleon mass, defined in (\ref{mainresult}) and (\ref{mainresult2})
respectively. By a single star we mark the coefficients obtained by the simplified calculation (by reducing the higher powers of axial coupling
constant). By a double star we mark the coefficients obtained by evaluation of the conjecture (\ref{6:M(m)}).} \label{table_mass}
\end{table}

The calculation of the even coefficients $k$ can be significantly simplified by using the conjectures discussed below. So, by neglecting higher
powers of $g_A$ during the evaluation of the diagrams, we could also evaluate the five-loop coefficient $k_{12}$. Adding the further conjecture
about the relation with the LLog in the pion mass, we can obtain the six and seven-loop coefficients $k_{14}$ and $k_{16}$. However, these
coefficients are the result of conjectures and, therefore, in the table 1 they are marked by stars.

\section{Discussion}

The obtained result of the straightforward calculation, i.e. the coefficients $k_1,\ldots,k_{11}$ shows a number of regularities. Some of the
regularities we can explain easily, while some of them we cannot. But we are sure that this regularities are not accidental and are an example
of the deep connection of the LLog approximation (in non-renormalizable theories) with the structure of theory.

The most intriguing observation is that the coefficients $k_{2n}$ contains a very particular combination of LECs. The pattern appearing in
$k_{2n}$ is not well understood yet. While it is clear why the coupling constant $c_4$ does not participate in the nucleon mass at LLog (
because it produces an $\epsilon_{\mu\nu\alpha\beta}$, which is P-odd), we have not found any simple argument why $g_A$ only appears up to order
$g_A^2$, and what is the special in the combination $(g_A^2+M(c_2+4c_3-4c_1))$.

From the general point of view, one expects that the LLog coefficient should be linear in $c_i$, but a general polynomial in $g_A$. Because the
number of vertices from $\mathcal{L}^{(1)}$ is restricted to one (see fig.2 and explanations in the previous section), but the number of
vertices from $\mathcal{L}^{(0)}$ is naturally unrestricted. Moreover, the expression for the propagator contains all allowed powers of $g_A$.
These powers cancel in the solution of the pole equation on the physical mass. The deeper consideration of these cancelation (see discussion in
\cite{Bijnens:2014ila}) hints that that absence of higher powers of $g_A$ in coefficients $k_{2n}$ is a consequence of Lorentz invariance, and
of the additional subtractions of infrared (heavy mass) singularities into renormalization counter terms within HBChPT. However, we have not
been able to prove this. Supposing that the cancelation of the higher powers of $g_A$ takes place at all orders, one can neglect these powers
during the computation of diagrams. This procedure significantly reduces the demands for computer time and allows us to calculate the
coefficient $k_{12}$.

Even more intriguing result reveals if one inverts the perturbation series and presents the nucleon mass LLog coefficient in terms of the
physical pion mass $m_{\text{phys}}$
\begin{eqnarray}
\label{mainresult2} M_{\text{phys}}&=& M+\frac{m_\text{phys}^2}{M} \sum_{n=1}^\infty r_{2n} L_\pi^{n-1}
 +\pi m_\text{phys}\frac{m_\text{phys}^2}{(4\pi F)^2}
     \sum_{n=1}^\infty r_{2n+1} L_\pi^{n-1},
\end{eqnarray}
where
$$
L_\pi=\frac{m^2_\text{phys}}{(4\pi F)^2}\log\(\frac{\mu^2}{m_\text{phys}^2}\).
$$
The coefficients $r_n$ of this expansion are presented in table 1.

\begin{figure}[tb!]
\centering
\includegraphics[height=.25\textheight]{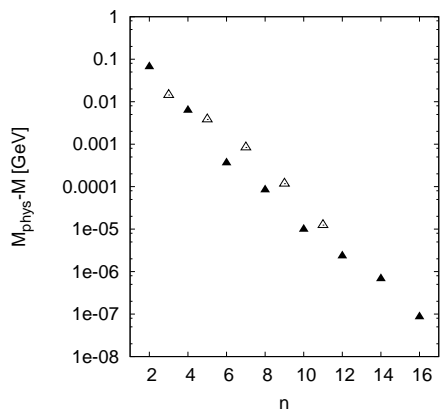}
\caption{The absolute value of the contribution of the individual terms ($\sim k_n$) in (\ref{mainresult}) at $m=138$~MeV. Open symbols are the
odd order coefficients. Filled symbols are the even order order coefficients.} \label{figparts}
\end{figure}

One can see that the non-analytical in quark mass terms $r_\text{odd}$ do not simplify in this form of expansion, while the expressions for the
coefficient $r_\text{even}$ are significantly simplified. Moreover, the combination of LECs $(g_A^2+M(c_2+4c_3-4c_1))$ completely disappears
from the higher order terms. The rest coefficient precisely repeat the LLog expansion for $m^4_\text{phys}(\mu')$. Thus, we conjecture the LLog
part of the expression for the nucleon bare mass via the physical masses \textit{at all orders} to be
\begin{eqnarray}\label{6:M(m)}
M= M_\text{phys} +\frac{3}{4}m_\text{phys}^4 \frac{\log\(\frac{\mu^2}{m_\text{phys}^2}\)}{(4\pi F)^2}
\(\frac{g_A^2}{M_\text{phys}}-4c_1+c_2+4c_3\) \qquad\qquad
\nn\\
 -\frac{3c_1}{{(4\pi F)^2}}\int\limits_{m^2_\text{phys}}^{\mu^2}m_\text{phys}^4(\mu')~\frac{d\mu'^2}{\mu'^2}.
\end{eqnarray}
The expression for the physical pion mass is known up to 6-loop order (\ref{2:mPhys}), therefore, we can obtain two more LLog coefficients for
the physical nucleon mass. These are presented in the table 1 and indicated by the double-star marks.

As we mentioned earlier, the LLog are not necessarily dominant. They do however give an indication of the size of corrections to be expected. To
show an example, we plot in fig.3 the absolute value of the individual terms containing $k_i$ of (\ref{mainresult}) for $m=138$~MeV. Note the
excellent convergence. We have used here one set of inputs for the $c_i$ as determined in \cite{Bernard:1995dp} and reasonable values for the
other quantities.

\begin{theacknowledgments}
This work is supported in part by the European Community-Research Infrastructure Integrating Activity Study of Strongly Interacting Matter"
(HadronPhysics3, Grant Agreement No. 28 3286) and the Swedish Research Council grants 621-2011-5080 and 621-2013-4287.
\end{theacknowledgments}

\bibliographystyle{aipproc}

\end{document}